\documentclass[ reprint,superscriptaddress, amsmath,amssymb,aps]{revtex4-2}
\usepackage{geometry}                
\usepackage{graphicx}
\usepackage{amssymb}
\usepackage{xcolor}

\graphicspath{ {./} }

\makeatletter
\renewcommand*{\@fnsymbol}[1]{\ensuremath{\ifcase#1\or *\or \mathrm{a}\or \mathrm{b}\or \mathrm{c}\or \mathrm{d}\or \mathrm{e}\or \dagger \else\@ctrerr\fi}}
\makeatother

\begin{document}

\title{Charging of free-falling test masses in orbit due to cosmic rays: results from LISA Pathfinder}
\def\addressa{European Space Astronomy Centre, European Space Agency, Villanueva de la Ca\~{n}ada, 28692 Madrid, Spain}
\def\addressb{Albert-Einstein-Institut, Max-Planck-Institut f\"ur Gravitationsphysik und Leibniz Universit\"at Hannover, Callinstra{\ss}e 38, 30167 Hannover, Germany}
\def\addressc{Universit\'e Paris Cit\'e, CNRS, Astroparticule et Cosmologie, F-75013 Paris, France}
\def\addressd{High Energy Physics Group, Physics Department, Imperial College London, Blackett Laboratory, Prince Consort Road, London, SW7 2BW, UK }
\def\addresse{Dipartimento di Fisica, Universit\`a di Roma ``Tor Vergata'',  and INFN, sezione Roma Tor Vergata, I-00133 Roma, Italy}
\def\addressf{Department of Industrial Engineering, University of Trento, via Sommarive 9, 38123 Trento, and Trento Institute for Fundamental Physics and Application / INFN}
\def\addressh{European Space Technology Centre, European Space Agency, Keplerlaan 1, 2200 AG Noordwijk, The Netherlands}
\def\addressi{Dipartimento di Fisica, Universit\`a di Trento and Trento Institute for Fundamental Physics and Application / INFN, 38123 Povo, Trento, Italy}
\def\addressj{The School of Physics and Astronomy, University of Birmingham, Birmingham, UK}
\def\addressl{Institut f\"ur Geophysik, ETH Z\"urich, Sonneggstrasse 5, CH-8092, Z\"urich, Switzerland}
\def\addressm{The UK Astronomy Technology Centre, Royal Observatory, Edinburgh, Blackford Hill, Edinburgh, EH9 3HJ, UK}
\def\addressn{Institut de Ci\`encies de l'Espai (ICE, CSIC), Campus UAB, Carrer de Can Magrans s/n, 08193 Cerdanyola del Vall\`es, Spain}
\def\addresso{DISPEA, Universit\`a di Urbino ``Carlo Bo'', Via S. Chiara, 27 61029 Urbino/INFN, Italy}
\def\addressp{European Space Operations Centre, European Space Agency, 64293 Darmstadt, Germany}
\def\addressq{Physik Institut, Universit\"at Z\"urich, Winterthurerstrasse 190, CH-8057 Z\"urich, Switzerland}
\def\addressr{SUPA, Institute for Gravitational Research, School of Physics and Astronomy, University of Glasgow, Glasgow, G12 8QQ, UK}
\def\addresss{Department d'Enginyeria Electr\`onica, Universitat Polit\`ecnica de Catalunya,  08034 Barcelona, Spain}
\def\addresst{Institut d'Estudis Espacials de Catalunya (IEEC), C/ Gran Capit\`a 2-4, 08034 Barcelona, Spain}
\def\addressu{Gravitational Astrophysics Lab, NASA Goddard Space Flight Center, 8800 Greenbelt Road, Greenbelt, MD 20771 USA}
\def\addressbb{Department of Mechanical and Aerospace Engineering, MAE-A, P.O. Box 116250, University of Florida, Gainesville, Florida 32611, USA}
\def\addresscc{Istituto di Fotonica e Nanotecnologie, CNR-Fondazione Bruno Kessler, I-38123 Povo, Trento, Italy}
\def\addressdd{isardSAT SL, Marie Curie 8-14, 08042 Barcelona, Catalonia, Spain}
\def\addressee{Escuela Superior de Ingenier\'ia, Universidad de C\'adiz, 11519 C\'adiz, Spain}
\def\addressff{Department of Physics, P.O. Box 118440, University of Florida, Gainesville, Florida 32611, USA}
\def\addressgg{IRFU, CEA, Universit\'{e} Paris-Saclay, F-91191 Gif-sur-Yvette, France}
\def\addresshh{isardSAT SL, Marie Curie 8-14, 08042 Barcelona, Catalonia, Spain}
\def\addressii{Institut f\"ur Theoretische Physik, Universit\"at Heidelberg, Philosophenweg 16, 69120 Heidelberg, Germany}
\def\addressjj{Universidad Loyola Andaluc\'{i}a, Department of Quantitative Methods, Avenida de las Universidades s/n, 41704 Dos Hermanas, Sevilla, Spain.}
\def\addresskk{INAF Osservatorio Astronomico di Capodimonte, I-80131 Napoli, Italy}

\author{M.~Armano}\affiliation{\addressh}
\author{H.~Audley}\affiliation{\addressb}
\author{J.~Baird}\affiliation{\addressc}
\author{P.~Binetruy}\thanks{Deceased}\affiliation{\addressgg}
\author{M.~Born}\affiliation{\addressb}
\author{D.~Bortoluzzi}\affiliation{\addressf}
\author{E.~Castelli}\altaffiliation[Present address:~]{\addressu}\affiliation{\addressi}
\author{A.~Cavalleri}\affiliation{\addresscc}
\author{A.~Cesarini}\affiliation{\addresso}
\author{A.\,M~Cruise}\affiliation{\addressj}
\author{K.~Danzmann}\affiliation{\addressb}
\author{M.~de Deus Silva}\affiliation{\addressa}
\author{I.~Diepholz}\affiliation{\addressb}
\author{G.~Dixon}\affiliation{\addressj}
\author{R.~Dolesi}\affiliation{\addressi}
\author{L.~Ferraioli}\affiliation{\addressl}
\author{V.~Ferroni}\affiliation{\addressi}
\author{E.\,D.~Fitzsimons}\affiliation{\addressm}
\author{M.~Freschi}\affiliation{\addressa}
\author{L.~Gesa}\thanks{Deceased}\affiliation{\addressn}
\author{D.~Giardini}\affiliation{\addressl}
\author{F.~Gibert}\altaffiliation[Present address:~]{\addressdd}\affiliation{\addressi}
\author{R.~Giusteri}\affiliation{\addressb}
\author{C.~Grimani}\affiliation{\addresso}
\author{J.~Grzymisch}\affiliation{\addressh}
\author{I.~Harrison}\affiliation{\addressp}
\author{M.-S.~Hartig}\affiliation{\addressb}
\author{G.~Heinzel}\affiliation{\addressb}
\author{M.~Hewitson}\affiliation{\addressb}
\author{D.~Hollington}\affiliation{\addressd}
\author{D.~Hoyland}\affiliation{\addressj}
\author{M.~Hueller}\affiliation{\addressi}
\author{H.~Inchausp\'e}\altaffiliation[Present address:~]{\addressii}\affiliation{\addressc}
\author{O.~Jennrich}\affiliation{\addressh}
\author{P.~Jetzer}\affiliation{\addressq}
\author{N.~Karnesis}\affiliation{\addressc}
\author{B.~Kaune}\affiliation{\addressb}
\author{C.\,J.~Killow}\affiliation{\addressr}
\author{N.~Korsakova}\affiliation{\addressc}
\author{J.\,A.~Lobo}\thanks{Deceased}\affiliation{\addressn}
\author{J.\,P.~L\'{o}pez-Zaragoza}\affiliation{\addressn}
\author{R.~Maarschalkerweerd}\affiliation{\addressp}
\author{D.~Mance}\affiliation{\addressl}
\author{V.~Mart\'{i}n}\affiliation{\addressn}
\author{J.~Martino}\affiliation{\addressc}
\author{L.~Martin-Polo}\affiliation{\addressa}
\author{F.~Martin-Porqueras}\affiliation{\addressa}
\author{P.\,W.~McNamara}\affiliation{\addressh}
\author{J.~Mendes}\affiliation{\addressp}
\author{L.~Mendes}\affiliation{\addressa}
\author{N.~Meshksar}\affiliation{\addressl}
\author{M.~Nofrarias}\affiliation{\addressn}
\author{S.~Paczkowski}\affiliation{\addressb}
\author{M.~Perreur-Lloyd}\affiliation{\addressr}
\author{A.~Petiteau}\affiliation{\addressc}\affiliation{\addressgg}
\author{E.~Plagnol}\affiliation{\addressc}
\author{J.~Ramos-Castro}\affiliation{\addresss}
\author{J.~Reiche}\affiliation{\addressb}
\author{F.~Rivas}\altaffiliation[Present address:~]{\addressjj}\affiliation{\addressn}
\author{D.\,I.~Robertson}\affiliation{\addressr}
\author{G.~Russano}\altaffiliation[Present address:~]{\addresskk}\affiliation{\addressi}
\author{J.~Slutsky}\affiliation{\addressu}
\author{C.\,F.~Sopuerta}\affiliation{\addressn}
\author{T.\,J.~Sumner}\affiliation{\addressd}\affiliation{\addressff}
\author{D.~Texier}\affiliation{\addressa}
\author{J.\,I.~Thorpe}\affiliation{\addressu}
\author{D.~Vetrugno}\affiliation{\addressi}
\author{S.~Vitale}\affiliation{\addressi}
\author{G.~Wanner}\affiliation{\addressb}
\author{H.~Ward}\affiliation{\addressr}
\author{P.\,J.~Wass}\email[Corresponding author.~]{pwass@ufl.edu}\affiliation{\addressd}\affiliation{\addressbb}
\author{W.\,J.~Weber}\affiliation{\addressi}
\author{L.~Wissel}\affiliation{\addressb}
\author{A.~Wittchen}\affiliation{\addressb}
\author{P.~Zweifel}\affiliation{\addressl}

\date{draft of \today}                                           


\begin{abstract}

A comprehensive summary of the measurements made to characterize test mass charging due to the space environment during the LISA Pathfinder mission is presented. 
Measurements of the residual charge of the test mass after release by the grabbing and positioning mechanism, show that the initial charge of the test masses was negative after all releases, leaving the test mass with a potential in the range $-12$\,mV to $-512$\,mV.
Variations in the neutral test mass charging rate between 21.7 and 30.7\,e\,s$^{-1}$ were observed over the course of the 17-month science operations produced by cosmic ray flux changes including a Forbush decrease associated with a small solar energetic particle event. 
A dependence of the cosmic ray charging rate on the test mass potential between $-30.2$ and $-40.3$\,e\,s$^{-1}$\,V$^{-1}$ was observed resulting in an equilibrium test mass potential between 670 and 960\,mV, and this is attributed to a contribution to charging from low-energy electrons emitted from the gold surfaces of the gravitational reference sensor. 
Data from the on-board particle detector show a reliable correlation with the charging rate and with other environmental monitors of the cosmic ray flux. This correlation is exploited to extrapolate test mass charging rates to a 20-year period giving useful insight into the expected range of charging rate that may be observed in the LISA mission.
\end{abstract}
\maketitle


Free-falling test masses in precision experiments are susceptible to parasitic electrostatic forces and torques \cite{jafry1996}, with significant contributions from the interaction between stray electric fields and any electrostatic charge~\cite{Antonucci, Armano2017a,Sumner2019}. These effects can make up a sizeable fraction of the noise budget for future gravitational wave observatories such as the Laser Interferometer Space Antenna (LISA)~\cite{LISA2017} which will have a measurement bandwidth between 0.1\,mHz and 1\,Hz. LISA Pathfinder (LPF)~\cite{McNamara2008} was a European Space Agency mission to demonstrate technology for LISA, especially in measuring the low-frequency acceleration noise acting on the test masses coming from the spacecraft environment \cite{Armano2016,Armano2018c}. 

In this paper we present measurements of the test-mass charging due to the space environment over the course of the LISA Pathfinder mission and discuss the observed variations under a number of different operational conditions.
The paper is organized as follows: in Section \ref{Sec:intro} we review the level of charging due to cosmic rays predicted before the launch of LPF. We describe the hardware on board relevant to measuring the test-mass charge and monitoring the space environment responsible for charging.
In Section~\ref{Sec:lpf_charge_campaign} we describe the method for evaluating the test mass charge on LPF. We present an overview of the charge history of the test masses during the mission including the remnant charge of the test mass after release by the grabbing and positioning mechanism.
In Section~\ref{Sec:electrostatics} evidence of the variation in test-mass charging caused by electrostatic effects in the LPF Gravitational Reference Sensor are presented. 
Section~\ref{Sec:RateVsFlux} describes the observed dependence of the charging rate on the cosmic ray flux and correlation with the on board radiation monitor. 
Section~\ref{Sec:LISA_timeline} presents an extrapolation of the LPF charging rate across a full solar cycle and finally in Section~\ref{discussion} we present our conclusions.

\section{Introduction}
\label{Sec:intro}

\subsection{Charge accumulation and discharge}
\label{intro:Q_accum}

The test mass charge accumulates due to the energetic, charged-particle space environment, dominated by protons and $^4$He. These particles which may be galactic cosmic rays (GCR) or solar energetic particles (SEP) are able to penetrate the shielding of the spacecraft and deposit charge on the test masses if they have an energy above  $\sim$100\,MeV\,nucl$^{-1}$~\cite{jafry1996}. 

Prior to launch, the expected charge rate from both cosmic rays and solar energetic particles had been evaluated using a number of simulations \cite{Araujo2005,Wass2005, Grimani2005,Grimani2005b,Grimani2015} using G\textsc{eant}4~\cite{geant4} and FLUKA~\cite{fluka} high energy physics simulation toolkits.
Predictions of the mean charging rate were between +38 and +44\,e\,s$^{-1}$ (where e is the proton charge) in solar minimum galactic cosmic ray conditions falling to roughly half that during solar maximum. LISA Pathfinder began science operations in March 2016 as the solar cycle was approaching solar minimum conditions. 

The mean charging rate is the result of both negative and positive charging with a range of multiplicity events of both polarities. 
Hence the predicted stochastic noise behaviour in the charging rate was higher than expected from the mean rate. The noise is characterised by an effective rate of single charges, $\lambda_{\mathrm{eff}}$, that would create the observed noise. Simulations predicted a value for $\lambda_{\mathrm{eff}}\sim200$-$400$\,s$^{-1}$.  

The first in-orbit observations of the LPF test-mass charging rate and its stochastic noise based on a measurement made on 2016-04-20 were presented in \cite{Armano2017b}. Charging rates were found to be 22.9-24.5\,e\,s$^{-1}$ and $\lambda_{\mathrm{eff}}=1100$-$1400$\,s$^{-1}$
The observation of a significantly higher charge noise, together with the charge rate measurements presented here have motivated a continuing effort to characterize in detail the physics of the charging process~\cite{grimani21a, grimani21b, grimani22,Taioli2023,Wass2022}. In particular, reported for the first time here is experimental evidence that a stable equilibrium for the test mass charging due to the space environment can be reached. This equilibrium can be explained by a population of low-energy (eV) electrons created by secondary emission at the TM and EH surfaces and the electric fields within the GRS~\cite{Wass2022}.


\subsection{The Gravitational Reference Sensor}
\label{intro:GRS}

\begin{figure}
\centering
\includegraphics[scale=0.1, trim=4cm 0.2cm 0.2cm 0.2cm, clip]{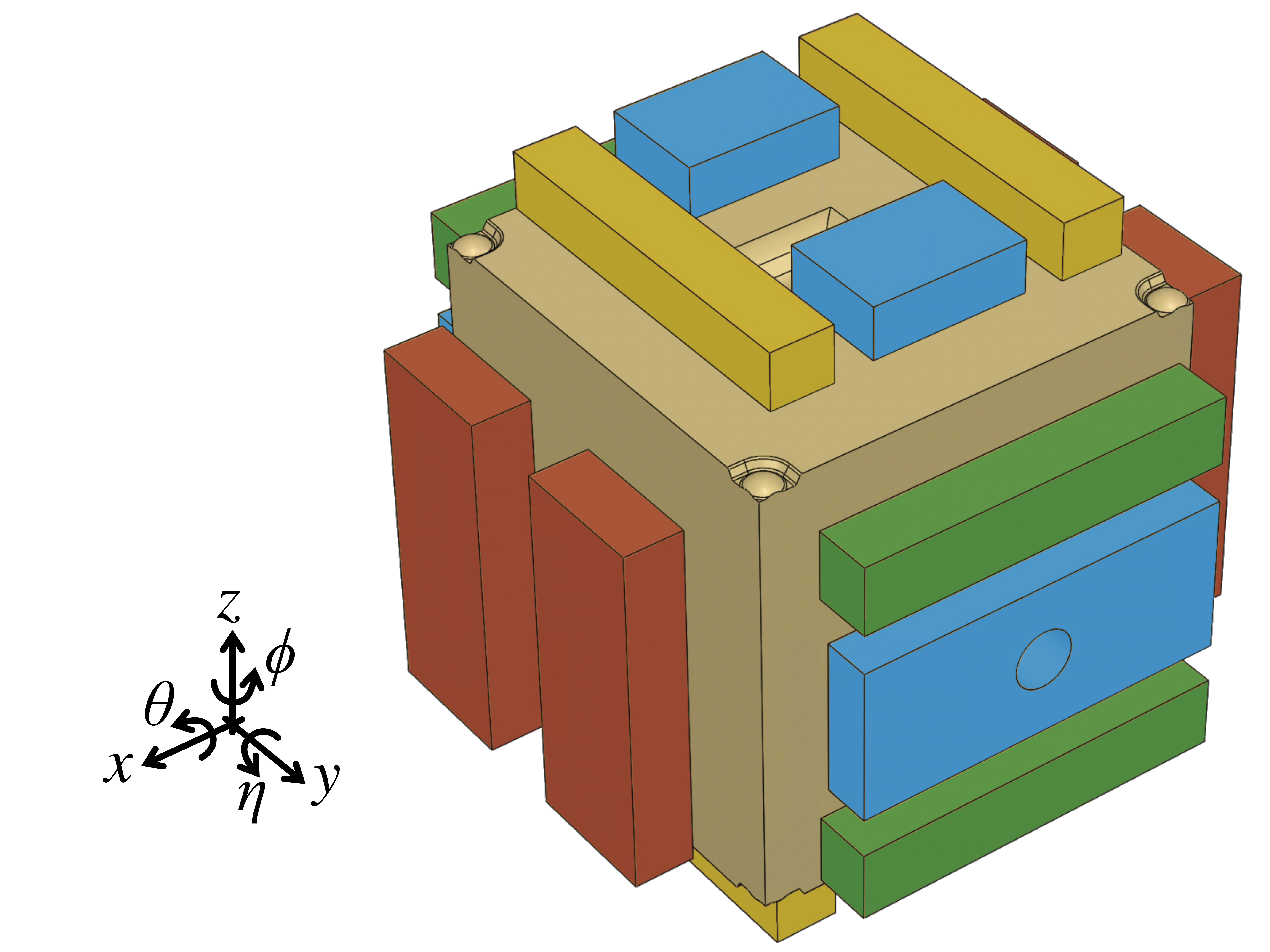}
\caption{\label{fig:GRS}Diagram of LISA Pathfinder GRS showing 46\,mm cubic test mass surrounded by capacitive-sensing bias injection electrodes in blue, $x$/$\phi$ sensing and actuation electrodes in red, $y$/$\theta$  in green and $z$/$\eta$ in yellow.}
\end{figure}

The LPF payload consisted of two Gravitational Reference Sensors (GRS), each comprised of a $\sim$2\,kg gold-plated Au/Pt test mass surrounded by a Au-plated electrode housing (EH) consisting of 18 electrodes as illustrated in Fig.~\ref{fig:GRS}. Twelve electrodes separated from the test mass by gaps of 2.9--4.0\,mm were used for position sensing and to apply actuation force and torques to the test mass~\cite{Armano2017b}. Actuation voltages were applied at ac audio frequencies nominally balanced around the ground potential of the GRS front-end electronics (FEE).
Slowly varying or dc voltages could also be applied as described below in order to measure test mass charge. A further six electrodes on the $y$ and $z$ axes of the test mass carried a 4.88\,V amplitude 100\,kHz ac voltage which induced a 0.6\,V 100\,kHz test mass potential bias used for capacitive position sensing. The test mass was completely isolated from its surroundings with no electrical connection to ground.   
In order to reduce disturbances from electrostatic forces, the test mass charge had to be kept near zero. In LPF, this was achieved by illuminating the test mass and sensor with UV light  at 253.7\,nm from a low-pressure mercury discharge lamp. The resulting photoelectrons transfer charge to or from the test mass as required. The detailed behaviour of the discharging system is discussed in \cite{Armano2018a}. 

\subsection{Electrostatic forces}
\label{intro:EF}

The test mass charge was evaluated through its interaction with the electric fields in the GRS. In LPF, the most sensitive axis for force detection is the $x$-axis connecting the two test masses. The electrostatic force acting on the test mass (TM) in the $x$ axis is~\cite{jafry1996} 
\begin{equation}
	\label{eqn1} 
	F_x=\frac{1}{2}\sum\frac{\partial{C_i}}{\partial{x}} \left(V_i - V_{\textsc{tm}}\right)^2\,,
\end{equation}
where the sum is defined over all $i$ equipotential surface elements of the sensor with potential $V_i$ relative to ground and capacitance with respect to the test mass, $C_i$. All test mass surfaces surfaces are assumed to be at the same potential, $V_{\textsc{tm}}$, determined by the potential of all sensor surfaces and the free charge on the test mass $q$, according to
\begin{equation}
	\label{eqn2} 
	V_{\textsc{tm}} = \frac{\sum_i C_i V_i}{C_T} + \frac{q}{C_T}\,,
\end{equation}
where $C_T$ is the total capacitance of the test mass with respect to the surrounding surfaces (34.2\,pF for each test mass on LPF). \\
Except for dedicated characterization experiments, voltages applied to electrodes are chosen to give no in-band fluctuation of $V_{\textsc{tm}}$; however, unavoidable stray fields and in-band fluctuations of the potentials of the TM and surrounding surfaces give rise to forces that can, if left unchecked, compromise the acceleration noise performance of the instrument \cite{Armano2017a}. 

In order to measure the test mass potential, a force with frequency $f_{\textsc{mod}}$ is excited proportional to $V_{\textsc{tm}}$ according to Eq.~(\ref{eqn1}). Voltages $V_i=V_{\textsc{mod}} \sin 2\pi f_{\textsc{mod}} t$ are applied with opposite phase to the sensing electrodes on either side of the test mass which have equal and opposite capacitance derivatives $\frac{\partial{C_x}}{\partial{x}}$~(291\,pF\,m$^{-1}$). To maximize the precision of the measurement, $f_{\textsc{mod}}$ is chosen to in the sensitive, mHz band of the instrument. The resulting force along the $x$-axis, assuming all other potentials are zero, is:

\begin{equation}
	\label{eqn3} 
		F_x = -4 V_{\textsc{mod}} \frac{q}{C_{T}} \left| \frac{\partial{C_x}}{\partial{x}}\right| \sin{2\pi f_{\textsc{mod}}} \,.
\end{equation}
Equivalently, applying modulating voltages to the diagonally opposite electrodes produces a torque, $\Gamma$. For example in $\phi$ rotation:
\begin{equation}
	\label{eqn4} 
		\Gamma_{\phi} = -4 V_{\textsc{mod}} \frac{q}{C_T} \left| \frac{\partial{C_x}}{\partial{\phi}}\right|\sin{2\pi f_{\textsc{mod}}} \,.
\end{equation}

Mixing of the stray potential patches on the electrodes with the charge measurement modulation voltages will produce an additional force or torque term modulated at $f_{\textsc{mod}}$. This creates an unknown charge offset which, to first order is proportional to the average stray potential of the electrodes used to excite the charge measurement force or torque. 


\subsection{Radiation Monitor}
\label{intro:rm_lpf}
The radiation monitor (RM) on board LPF was specifically designed to monitor the interplanetary GCR and SEP environment responsible for charge build-up on the test masses~\cite{Wass2006}. Two large area (147\,mm$^2$), 300\,$\mu$m-thick silicon PIN photodiodes were arranged in a telescopic configuration in a copper enclosure to give shielding from particles with energy below roughly 100\,MeV\,nucl$^{-1}$. The deposited energy resolution was around 15\,keV which was sufficient to allow differentiation between cosmic ray particles---most of which are close to minimally ionizing in the silicon detectors---and SEPs which result in larger energy deposits. The intention was to use the RM to measure the dependence of TM charging on count rate in both GCR and SEP flux conditions. LPF experienced only one SEP event which did not produce a particle flux enhancement in the energy range relevant for charging. A Forbush decrease associated with that event was observed, however~\cite{Armano2018b}. The RM also produced a full history of the relative fluctuations in the the ambient cosmic ray flux with good counting statistics. With this data it has been possible to estimate the solar modulation parameter~\cite{Armano2018b}, and observe modulation of the cosmic ray flux due to solar magnetic field fluctuations \cite{Armano2018d}.


\section{Charge measurement campaign}
\label{Sec:lpf_charge_campaign}

\subsection{Experimental details}
\label{LPF:Q_measurement_process}

The test mass charge was measured on-board LPF using both force and torque excitations as described above. The charge on both test masses could be measured simultaneously using different modulation frequencies on each mass. The measurement precision depends on the amplitude of the applied voltages ($V_{\textsc{mod}}$) and the force (torque) sensitivity in the chosen degree of freedom. The most sensitive measurements were made using the on-board laser interferometer~\cite{Armano2021, Armano2022a} to measure their differential motion in either $x$ or $\phi$ (around $z$). 
The force on the test masses was recovered from their differential (angular) acceleration, accounting for control forces (torques), and in the case of the $x$-axis measurements, tilt-to-length and non-inertial effects due to spacecraft rotations.
Typically the sensitivity to the force (torque) excitation was limited by direct stray forces (torques) on the test mass. Along the $x$-axis, Brownian force noise on the test masses created a differential acceleration noise which decreased from around 10 to 3\,fm\,s$^{-2}\,\rm{Hz}^{-1/2}$ in the frequency range from 0.5 to 30 mHz over the course of the mission~\cite{Armano2018a}. In the $\phi$ rotation, the differential angular acceleration was typically 300\,frad\,s$^{-2}$\,Hz$^{-1/2}$ for $f_{\textsc{mod}}=5$\,mHz, comparable to the level expected from angular readout sensitivity.
The charge was calculated by applying  Eqs.~(\ref{eqn3}) or (\ref{eqn4}) 
to the force or torque time series, demodulating at $f_{\textsc{mod}}$. The residual accelerations mentioned above correspond to a TM charge (potential) measurement uncertainty of 0.24 and 0.77\,fC (7.0\,$\mu$V and 23$\mu$V) in a 200\,s measurement using $V_{\textsc{mod}}=50$\,mV on the $x$ and $\phi$ degrees of freedom respectively.

A systematic offset between test mass charge measurements made in the $x$ and $\phi$ degrees was observed during the mission. In a dedicated experiment to determine the magnitude of the effect, the test mass charge was measured sequentially with excitations along the two axes. The resulting potential difference between the two methods was $20.6\pm0.3$\,mV for TM\,1 and $11.6\pm0.5$\,mV for TM\,2 while the measurement of an applied polarization of the test mass was equal for both methods. 

The $x$ and $\phi$ charge measurement methods use the same electrodes with the same average stray potential to excite a force or torque on the test mass and therefore to first order, the offset in charge measurement should be equal. The difference observed could originate from the distribution of the patches over the electrode. In the $\phi$ measurement, patches would have a varying weighting in their contribution to the measurement offset according to the lever arm they exert on the test mass rotation.  The offset was measured only once during the LPF mission and therefore it is not possible to make definitive statement about its stability over time. However, another effect of voltage patches, charge to force-noise coupling has been found to be stable over the course of the mission \cite{Armano2017a}.

Extended measurements of the test mass charge allow its time dependence to be studied. The charging rate can be estimated by taking a linear fit to the time series. 
The uncertainty in the charge rate can be improved by increasing $V_{\textsc{mod}}$ or by measuring for longer. In practice, the latter is limited by low-frequency, $1/f$ noise arising from the charging shot noise accumulation and fluctuations in the particle flux responsible for charging. These have both been measured and reported in \cite{Armano2017a} and \cite{Armano2018b}, respectively. Typically, $V_{\textsc{mod}}$ was chosen to be relatively small (10-50\,mV) so as not to excite large signals in the instrument, and measurement durations were short compared to cosmic ray fluctuations. The dominating uncertainty in estimating the charge rate was therefore the test-mass acceleration noise limiting the charge measurement. For a charge measurement using the TM $x$-axis as described above, the charge-rate noise can be written as 
$S^{1/2}_R=m S^{1/2}_a(f_{\textsc{mod}}) C_T/ 4 V_{\textsc{mod}} \left| \frac{\partial C_x}{\partial x}\right| 2 \pi f_{\textsc{mod}}$ where $S^{1/2}_a(f_{\textsc{mod}})$ is the test mass acceleration noise spectral density at the modulation frequency and $m$ is the mass of the TM. Integrating over a 1-hr time interval results in a charge rate resolution of 0.34~e\,s$^{-1}$.

The LPF RM tracked variations in the particle flux at the spacecraft. For small changes in flux, the change in charge rate is expected to be proportional. Assuming a count rate of 9\,Counts\,s$^{-1}$, the error on the average count rate in a 1-hr time period is 0.05\,Counts\,s$^{-1}$. Results presented in Section~\ref{Sec:RateVsFlux} find a scale factor between charge rate and count rate of around 5\,e\,Count$^{-1}$ which when applied to the RM count rate resolution gives a corresponding charge rate estimate of 0.25\,e\,s$^{-1}$ over 1\,hour, comparable to the charge rate resolution using the $x$-axis with $V_{\textsc{mod}}=50$\,mV. Although the exact values depend on particle flux conditions, the RM is expected to be a useful diagnostic to confirm the origin of charge rate fluctuations observed during test mass measurements.


\subsection{Charge measurement results}

\begin{figure*}[htb]
\centering
\includegraphics[scale=0.8, trim=1.7cm 1.2cm 1.3cm 16cm, clip]{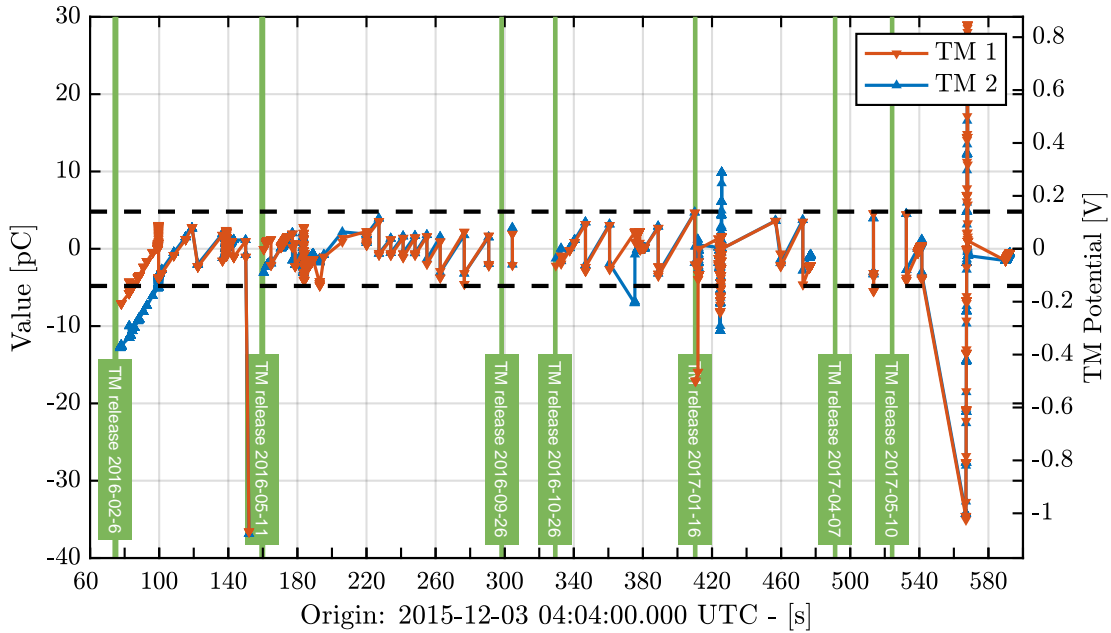}
\caption[LPF Q Meas.]{\label{fig:LPFQ}LPF test-mass charge as function of time throughout the mission. Dashed lines indicate the upper and lower limits for the charge during nominal operations at 3$\times10^7$\,e\,$\approx$\,140\,mV. Abrupt decreases in the test mass charge are the result of charge control operations using the onboard UV illumination system. Large excursions of the charge occurred at day 152, 425 and 567 during test mass charging and noise investigations.}
\end{figure*}

Charge measurements were made throughout the LPF mission as part of dedicated science investigations and as part of routine maintenance operations. A summary plot of measurements is shown in Fig.~\ref{fig:LPFQ}. 

In the early phase of the mission: days 80-120, the charge was measured regularly in order to understand the charging process and make the first discharging investigations. Over the following 100 days a number of further charge measurements were made aimed at characterizing the charge-related noise effects.

As indicated in Figure~\ref{fig:LPFQ} the test masses were regrabbed six times during the LISA Pathfinder science mission following the initial release in February 2016. In each case the test mass charge was measured soon afterwards and the test masses brought back into the operational charge range with the UV discharge system. In the last two releases, this process was automated on board LPF and the data necessary to recover the TM charge at release were not downlinked. In the first five releases however, the charge or potential $V_{\textsc{tm}}=q/C_T$ could be extrapolated to the point of release. Table \ref{Tab:Qrelease} lists the dates of the test mass releases and the TM potentials where known. The origin of the residual potential at release is the triboelectric charging of the test mass due to the release of the adhesion between the TM grabbing and positioning mechanism fingers and the test mass. A detailed analysis of this physical process is beyond the scope of this work but the observed potentials provide useful data for bounding the design parameters of the control loops for test mass capture during the LISA mission.

\begin{table}
\caption{TM potential after release}\label{Tab:Qrelease}
\begin{tabular}{c  r   r}
\hline
\hline
          & \multicolumn{2}{c}{$V_{\textsc{tm}}$ [mV]} \\
\cline{2-3}
Date  & TM\,1 & TM\,2 \\
    \hline
     2016-02-16    & $-421$ & $-281$\\ 
     2016-05-11    & $-98$  & $-52$\\ 
     2016-09-26    & $-98$ & $-80$\\ 
     2017-10-26    & $-61 $ & $-79$ \\ 
     2017-01-16    & $-12 $ & $-512$ \\ 
     2017-04-07    &   --     &  --\\
     2017-05-10    &    --    &  --\\
\hline
\end{tabular}
\end{table}

Later in the mission, the frequency of measurements was reduced but the charge was measured during regular station-keeping activities in order to maintain the test mass potential within the desired bounds. The charge was set using the discharge system to a negative value such that by the time of the next discharge, the charge would have accrued to the same value with opposite sign thus maximizing the time spent with the absolute charge below the required levels.  Station-keeping activities were carried out at 1, 2 and 3-week intervals at different stages of the mission. With the typical charge accumulation rate of around 25\,e\,s$^{-1}\approx10$\,mV\,day$^{-1}$, the target TM potentials for this periodic discharge activity varied between $-40$, $-80$ and $-120$\,mV. 

Other measurements with large TM charge excursions can be seen in the figure which were used to explore the properties of the TM charging process (as described here), charge-induced noise~\cite{Armano2017a} and the performance of the UV discharge system~\cite{Armano2018a}.


\section{Charge rate dependency on electrostatics}
\label{Sec:electrostatics}
Early test mass charge modelling work identified the possibility of kinetic emission from Au surfaces producing a very low-energy (eV) electron population between the test mass and its surroundings~\cite{Araujo2005}. It was postulated that this population could play a significant role in TM charging. Altering the test mass charge by photoelectric discharge changes the potential difference between the test mass and its surroundings through Eq.\,(\ref{eqn2}) providing a mechanism to probe the low energy electron population. 
Two experiments were conducted during the LPF mission dedicated to investigating this effect. In both cases, the TM charge was increased in a number of steps using the on-board UV charge control system. The test-mass charge was measured continuously, allowing an extended period between illuminations in which to evaluate the charge rate as a function of the average TM potential. 

The details for the two  measurements are shown in Table~\ref{tab:qv_measures}; during the first, the test mass potential was varied over a narrower range (around $-0.3$ to $+0.3$\,V) while in the later measurement, a broader range was explored ($-1$ to $+1$\,V). In each measurement $V_{\textsc{mod}}=50$\,mV and $f_{{\mathrm{MOD}}}=10$\,mHz for TM\,1 and 12\,mHz for TM\,2.

\begin{table*}
\caption{Measurement campaigns to investigate the charging rate dependence on test-mass potential}
\begin{tabular}{c r r r r}
\hline
\hline
    & \multicolumn{2}{c }{Meas.\,1} & \multicolumn{2}{c}{Meas.\,2}\\
    \hline
    Start date &  \multicolumn{2}{c }{2017-01-30} & \multicolumn{2}{c}{2017-06-22}\\
    Start time &  \multicolumn{2}{c }{19:00:00 UTC} & \multicolumn{2}{c}{02:12:00 UTC} \\
    Duration & \multicolumn{2}{c }{93600\,s} & \multicolumn{2}{c}{92550\,s} \\
    Number steps & \multicolumn{2}{c }{11} & \multicolumn{2}{c}{10} \\
     & \multicolumn{1}{c}{TM\,1} & \multicolumn{1}{c}{TM\,2} & \multicolumn{1}{c}{TM\,1}  & \multicolumn{1}{c}{TM\,2} \\
    \cline{2-5}
    Start potential [V] & $-0.310$ & $-0.236$ & $-1.001$  & $-1.011$ \\
    End potential [V] & $0.288$ & $0.046$ & $0.812$ & $0.857$\\
    Slope [e\,s$^{-1}$\,V$^{-1}$]& $-37.8\pm0.8$ & $-40.3\pm0.9$ & $-31.2\pm0.6$ & $-30.2\pm0.6$  \\
    $\frac{dq}{dt}$ at $V_{\textsc{tm}}=0$\,V [e\,s$^{-1}$] & $25.51\pm0.7$ & $27.5\pm0.1$ & $28.1\pm0.3$ & $28.9\pm0.6$ \\
    $V_{\textsc{tm}}$ at $\frac{dq}{dt}=0$ [mV] & $674\pm16$ & $682\pm16$ & $901\pm27$ & $957\pm39$\\
    Time constant [d] & 65 & 61 & 79 & 82\\
\hline
\hline
\end{tabular}
\label{tab:qv_measures}
\end{table*}

\begin{figure*}
\centering
\includegraphics[scale=0.8, trim=1.7cm 1.5cm 1.3cm 15.5cm, clip]{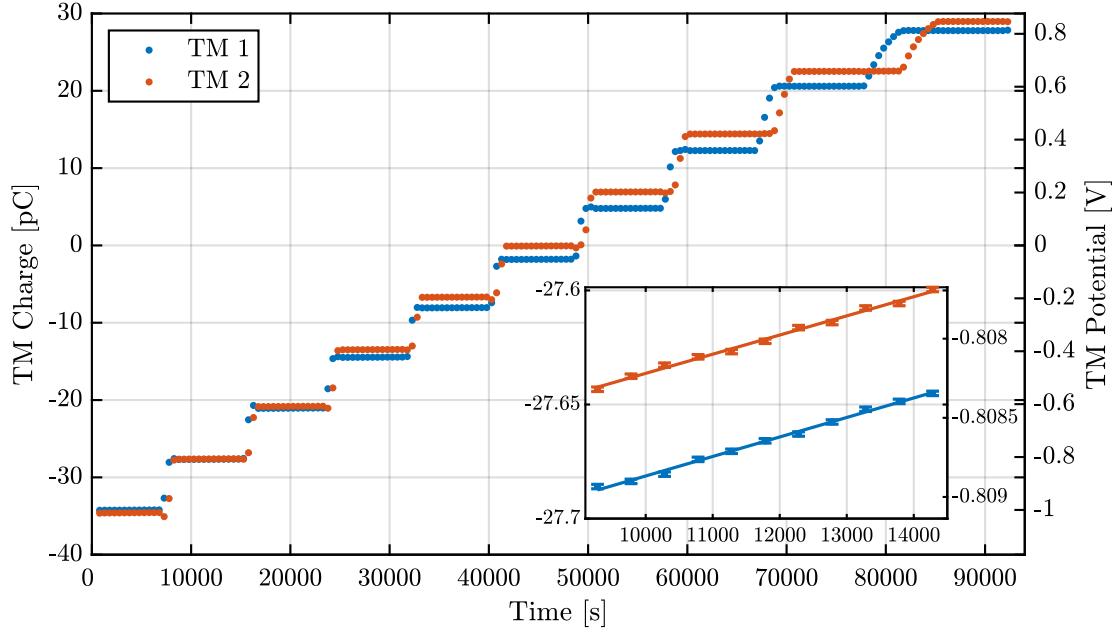}
\caption[LPF Qrate TM1 Meas.]{\label{fig:LPFQrate1} 
{LPF charging rate measurement for both test masses with minimum force and torque actuation authority and $V_{\textsc{mod}}=50$\,mV starting at 2017-06-22 at 02:12:00 UTC, the left-hand axis indicates the test mass charge and the right-hand axis the equivalent test mass potential with $C_{T}=34.1$\,pF. The inset shows an expanded plot of the data around   $V_{\textsc{tm}} \approx -808$\,mV with linear best fit lines. The charging rates from the fitted lines are $+53.4\pm1.2$\,e\,s$^{-1}$ ($\chi^2_{\mathrm{red}}=0.99$) and $52.4\pm1.2$\,e\,s$^{-1}$ ($\chi^2_{\mathrm{red}}$=1.2) for TM 1 and 2 respectively. }
}
\end{figure*}

Figure~\ref{fig:LPFQrate1} shows the test-mass charge throughout the second measurement on 2017-06-22. Each data point represents a charge measurement integrated over 500\,s. The errorbars shown are calculated from the estimated test mass acceleration noise and level of $V_{\textsc{mod}}$ using Eq.~(\ref{eqn3}), giving an error of $\pm$25\,fC. 
The duration of the illumination required to achieve charge steps increased as the test mass potential increased towards the equilibrium potential under UV illumination. 
The inset shows an enlargement of one of the charge plateaus, chosen when the charges of the two test masses were approximately equal at $-27.65$\,pC. The charge rate is determined with a relative precision of around 2\% in 4500\,s and the goodness of fit indicates that this precision is consistent with the measurement error.

\begin{figure}[htb]
\centering
\includegraphics[scale=0.48, trim=3cm 3cm 4cm 14cm, clip]{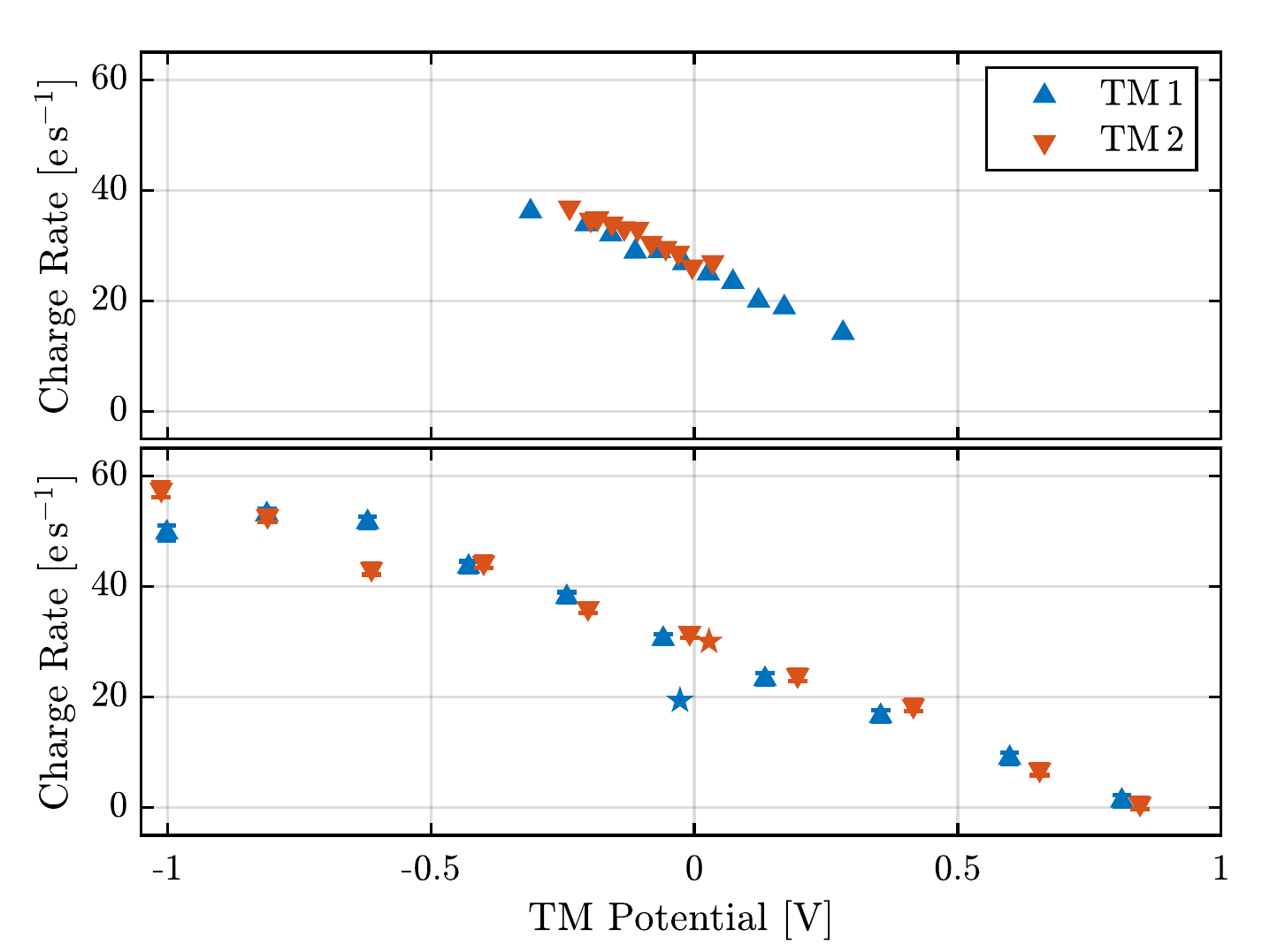}
\caption[Qrate vs TMv.]{\label{fig:Qrate_TMv} The variation in the test mass charging rates observed with varying test mass potential for the two measurements described in the text. The upper panel shows the results from 2017-01-30 and the lower panel form 2017-06-22.}
\end{figure}

Fig.~\ref{fig:Qrate_TMv} shows the evaluation of the charging rate at each step of test mass potential on test masses 1 and 2 during both experiments. In both cases, the environmental charging rate of both test masses show similar, approximately linear dependencies on $V_{\textsc{tm}}$.
Fitting a line to the data, we obtain the results shown in Table~\ref{tab:qv_measures}. The steep negative slopes immediately confirms that low-energy electrons are playing a significant role and the symmetry either side of $V_{\textsc{tm}}=0$\,V implies that these are originating from both the TMs and from the surrounding surfaces.  The data show that for $V_{\textsc{tm}}\approx+1$\,V the negative low-energy electron charging balances the positive charging from the high-energy cosmic rays and the overall charging rate goes to zero. This result is important as it would ensure that the system will reach an equilibrium in the case of a failure of the charge management system. The time constant of the approach to equilibrium---inversely proportional to the slope of the charge rate dependence on test mass potential---can be calculated from the slopes reported above. We estimate values of 61 to 82 days. Although the existence of an equilibrium is confirmed, the observed variation indicates that its value is uncertain and is unlikely to be accurately predictable for LISA. 

After the second scan, an additional single charge rate measurement was made on both test masses. On TM\,1 the charge was set so that $q/C_{T} = +0.9$\,V, dc voltages of $-4.8$\,V were applied to the $y$ and $z$ electrodes to bias the mass back to a neutral potential according to Eq.~(\ref{eqn2}). On TM\,2, the charge was set close to zero and no dc voltages were applied. The aim of this measurement was to probe the energy distribution of low-energy electrons contributing to charging at a higher energy range---in the regime of potential differences up to 4.8\,V---than could be reached by varying the test mass potential alone.
In this measurement, the expectation was that the transfer of kinetic electrons from TM\,1 to the facing $y$ and $z$ electrodes would be suppressed, but not at all suppressed over the rest of the surface of the test mass. This should lead to an overall reduction in the charging rate compared to a measurement with the test mass potential at 0\,V but no applied dc electrode voltages. A detailed analysis of this measurement relies on modeling of electron emission and the electrostatic fields within the EH~\cite{Wass2022}. 

The additional data points near 0\,V in Fig.~\ref{fig:Qrate_TMv} show the results of this measurement. Under these conditions with the overall TM\,1 potential at $-0.03$\,V the test mass charging rate was 19.3\,e\,s$^{-1}$. At the same time, the rate measured on TM 2 with no applied voltages and $V_{\textsc{tm}}=+0.03$\,V was 30.2\,e\,s$^{-1}$.

\begin{figure}[htb]
\centering
\includegraphics[scale=0.48, trim=2.9cm 3cm 4cm 14cm, clip]{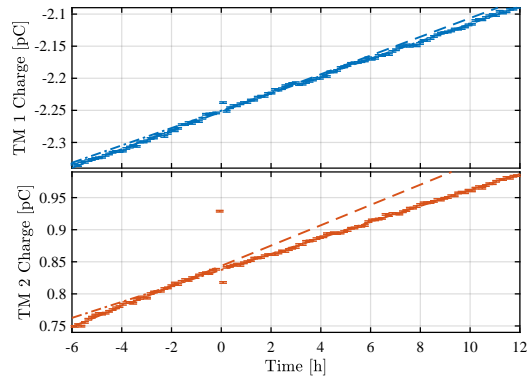}
\caption[Qrate vs Act.]{\label{fig:Qrate_force} The test mass charge measured with $V_{\textsc{mod}}=50$\,mV during the change in test mass force and torque authority at 2016-05-13 07:30 UTC ($t=0$). Dashed lines indicate the best fit to the charge trend at $t<0$ in high-authority mode and dash-dotted lines show the trend at $t>0$ in miminum-authority mode.}\label{Fig:TMChargingACVoltages}
\end{figure}

Additional evidence for the influence of low-energy electrons on the charging rate comes from a charge measurement made during 2016-05-13 as the instrument was recovering from a test mass release.
The charge rate was measured on both test masses with $V_{\textsc{mod}}=50$\,mV from 2016-05-13 00:00:00 UTC. Initially the test mass actuation system was in a high-authority mode with higher ac control voltages applied. Each of the 12 actuation electrodes is applied with a sum of ac voltages which control the one force and one torque degree of freedom, except the $x$ electrodes of TM~1 which control only torque. In high-authority mode, the typical peak amplitudes (the sum of two sinusoids) were around 6\,V and 9\,V on the $y$ and $z$ electrodes, 2\,V on the TM~1 $x$ electrodes and 5\,V on TM~2 $x$. At 2016-05-13 07:30:00 UTC, the control authority was decreased to optimize force noise performance~\cite{Armano2018c}, resulting in lower ac voltage amplitudes on all electrodes (around 3\,V on the $y$ and $z$ electrodes, 0.5\,V on the TM~1 $x$ electrode and 1\,V on TM~2 $x$). 
Figure~\ref{Fig:TMChargingACVoltages} shows the test mass charge measurements before and after the actuation change at $t$=0. A change in slope can be observed in the build up of charge on both test masses. Fitting to the data, we find the TM~1 rate goes from 25.3$\pm$0.1\,e\,s$^{-1}$ to 23.45$\pm$0.05\,e\,s$^{-1}$ and the TM~2 rate changes from 27.4$\pm$0.1 to 21.59$\pm$0.05\,e\,s$^{-1}$. 

The observed dependence of test-mass charging on ac voltages amplitude are likely related to the low-energy electron population. The observed behavior would be consistent with a increase in the net flow of electrons from the EH to the TM. This may be caused by an asymmetry in the electron emission from surfaces where ac voltages are applied, or electrostatic effects such as electric field edge effects which could alter the charge transfer between EH and TM as the magnitudes of the applied voltages change.  

\section{Charge rate vs GCR flux}
\label{Sec:RateVsFlux}


\subsection{Measurements dispersed through the mission}

\begin{figure}[htb]
\centering
\includegraphics[scale=0.48, trim=3cm 3cm 4cm 14cm, clip]{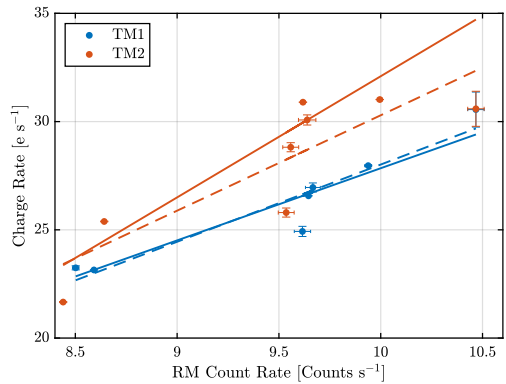}
\caption[LPF QRvsCR Meas.]{\label{fig:QRvsCRMeas}LPF test-mass charging rates as a function radiation monitor count rate for selected measurements with $V_{\textsc{tm}}$ close to 0\,V and ac actuation voltages in minimum-authority mode. Solid lines show the weighted best fit to the data, dashed lines indicate the unweighted fit.}
\end{figure}

Over the course of the mission the particle flux as measured by the in-situ particle detector increased by just less than 40\%~\cite{Armano2018b}. Selecting charge rate measurements done under similar bias conditions, to avoid complications caused by the behavior of the low-energy electron population,  the results can be plotted as a function of the radiation monitor count rate and compared to the expectation from the previous section. In Figure~\ref{fig:QRvsCRMeas} we have selected charge rate measurements in which the magnitude of the test mass charge was less than 2\,pC (V$_{\textsc{tm}}<60$\,mV) and the test masses actuation schemes were in a low force authority mode (as in the second half of the measurement shown in Figure \ref{Fig:TMChargingACVoltages}).  
While the physical processes governing the two measurements are rather different---the TM charging per particle is energy-dependent while the RM count rate is not---over the range of variation observed during the LPF mission, there is a clear correlation between them. For this reason we do not expect a linear dependence to extend to a zero count rate.  
The lines on the figure show weighted linear fits to the data. 
The resulting slopes are $3.3$\,e\,Count$^{-1}$ and $5.6$\,e\,Count$^{-1}$ for TM~1 and TM~2 respectively, however it is clear that the error bars are not consistent with a linear dependence in this case. 
Assuming there are underlying systematic errors which are not reflected in the statistics of the individual measurements, we can repeat the fit unweighted and assume a reduced $\chi^2$ of 1. This give slopes of 3.6$\pm$0.5\,e\,Count$^{-1}$ and 4.4$\pm$1.1\,e\,Count$^{-1}$. 

\begin{figure}[htb]
 \centering
 \includegraphics[scale=0.48, trim=3cm 3cm 4cm 14cm, clip]{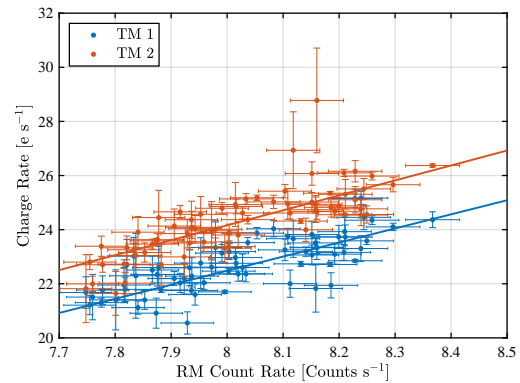}
 \caption[LPF QRvsCR 250.]{\label{fig:QRvsCR250}LPF test-mass charging rates plotted against radiation monitor count rate over a 3-day period beginning 2016-04-20 08:00:00 UTC. The measurement was performed with $V_{\textsc{mod}}=3$\,V and ac actuation voltages in high-authority mode but with reduced amplitudes on the $x$-electrodes. Charging and radiation monitor count rates are evaluated over 1-hour periods. Solid lines show the best fit to the data.}
 \end{figure}

Not included in Fig.~\ref{fig:QRvsCRMeas} because it was made with a higher test mass force authority, is the long charge measurement presented in Ref.~\cite{Armano2017b} which began at 2016-04-20 08:00:00 UTC. 
The ac actuation voltages in this case were similar to the high-authority mode described earlier but with reduced voltage amplitudes on the TM\,1 $x$ (1\,V) and TM\,2 $x$ (2\,V) electrodes.
This charge measurement has the highest modulation voltage of any measurements made during the mission at $V_{\textsc{mod}}=3$\,V and therefore the best precision on the charge rate determination. A measurable change in the radiation monitor count rate was also observed over the course of the 3-day measurement. Figure~\ref{fig:QRvsCR250} shows the charge rate plotted against the radiation count rate for both test masses. The charge rate data are calculated from the difference between consecutive charge measurements and averaged over 1-hour intervals, while the radiation monitor counts are averaged over the same time periods. The slopes to the best fit lines in this case were $5.2\pm0.3$\,e\,Count$^{-1}$ and $5.5\pm0.3$\,e\,Count$^{-1}$ for TM~1 and TM~2 respectively.

\subsection{A Forbush decrease}

\begin{figure}[htb]
\centering
\includegraphics[scale=0.48, trim=2.5cm 3cm 4.3cm 14cm, clip]{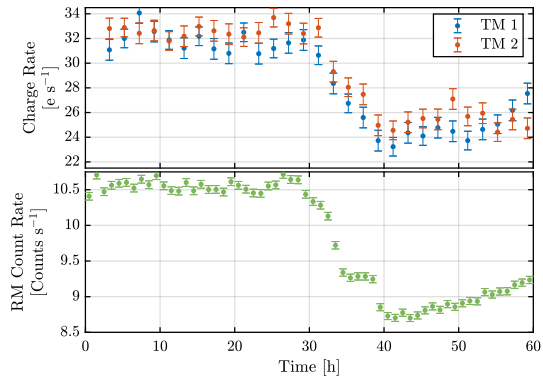}
\caption[Forbush Decrease.]{\label{fig:Forbush} Test mass charge and radiation monitor measurements made during the Forbush decrease beginning at 2017-07-15 01:00 UTC. The measurement was performed with $V_{\textsc{mod}}=10$\,mV and ac actuation voltages in minimum-authority mode. Charging and radiation monitor count rates are evaluated over 2-hour periods}
\end{figure}

\begin{figure}[htb]
\centering
\includegraphics[scale=0.48, trim=3cm 3cm 4cm 14cm, clip]{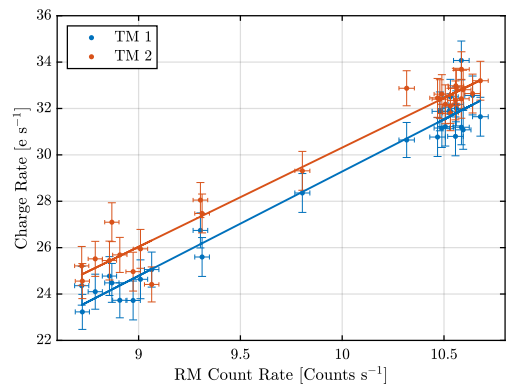}
\caption[Forbush Scatter.]{\label{fig:ForbushScatter} Scatter plot of the data shown in Figure~\ref{fig:Forbush} showing the correspondence between the test mass charging rate and radiation monitor count rate during the Forbush decrease. Solid lines show the best fit to the data.}
\end{figure}

On 2017-07-14 at approximately 03:00 UTC, a solar energetic particle event was recorded by a number of environmental monitors. Following this observation, a test mass charge measurement on both test masses with $V_{\textsc{mod}}=10$\,mV was commanded to begin onboard LPF, beginning at 2017-07-15 01:00:00 UTC. During this measurement the ac actuation voltages were in minimum-authority mode.
As reported in Refs.~\cite{Armano2018b,Armano2018d} the energy distribution of the solar protons was not sufficiently hard to register an increase in the onboard radiation monitor. However, following the passage of the solar particles, a temporary suppression of the galactic cosmic ray flux associated with a Forbush decrease was observed. 
Figure~\ref{fig:Forbush} shows the test-mass charging rates and radiation monitor count rate averaged over 2-hour time intervals during the event. Over the course of 12 hours from approximately 2017-07-16 08:00 UTC, the observed count rate decreased by around 17\%.  A scatter plot of the data is shown in Fig.~\ref{fig:ForbushScatter}.  The best-fit lines shown on the scatter plot have slopes of 4.5$\pm$0.2\,e\,Count$^{-1}$ and  4.3$\pm$0.2\,e\,Count$^{-1}$ for TM~1 and TM~2 respectively. 

Measurements presented in this section demonstrate that on timescales of days, changes in the average charging rate tracked the radiation monitor count rate with a linear relationship. The two measurements of the charge-count dependency made at the beginning and end of the mission disagree by several sigma in this dependency. However, both fall within the errors of the slope observed over the mission duration as whole shown in Figure~\ref{fig:QRvsCRMeas}.

\section{LISA timeline prediction}
\label{Sec:LISA_timeline}
Ref.~\cite{Armano2018b} established the close correspondence between in-situ measurements of the cosmic ray flux with the LPF radiation monitor and other solar activity indicators, especially the Integral Radiation Environment Monitor (IREM),~\cite{Hajdas2003b}. Using this correspondence, and the (assumed linear) correlation between test mass charging and the RM count rate,  we can produce timeline of test mass charging using historical data. 

The Level-0 IREM TC2 channel data is sensitive to protons in the same energy range as that responsible for test mass charging ($E>100$\,MeV\,nucl$^{-1}$)~\cite{Sandberg2012}. The available data time series extends from shortly after the launch of the INTEGRAL in October 2002 to the present day. The count rate data contain regular peaks as INTEGRAL traverses the Earth's radiation belts in its highly-elliptical, 2.7\,day orbit. These peaks are flagged in the data and have been removed during the analysis shown here. Enhancements in the data can also be seen due to solar energetic particle events. 
The correlation between count rate and charging is not expected to hold during SEP events given the difference in the energy spectrum of particles responsible for the flux enhancement. The energy distribution of GCR protons has a peak that varies from 400\,MeV to 800\,MeV during the solar cycle. In the range of energies relevant to test-mass charging however, SEP flux is typically distributed as a power law in energy. Due to the lower energies of the particles involved, the charging per incident particle is lower for SEPs. The RM count rate measurement on the other hand is insensitive to primary particle energy above around 100\,MeV.   
Despite this caveat, SEP events have not been removed from the timeline in order to give an impression of the frequency of the events. The scaled charge rate has, however, been truncated and so does not show the extrapolated charging rate during SEP events. 
The charging predictions for a LPF-like GRS are calculated from these data by scaling to the LISA Pathfinder count rate and then to TM charging. Over the course of the LPF mission, a fit to the RM singles count as a function of IREM TC2 count yields a scale factor of 2.8. From the previous section we adopt a factor of 4.4$\pm$1\,e\,Count$^{-1}$ to envelope all of the presented measurements. This charge rate dependence assumes the low-level of ac actuation voltages used for most of the LPF science mission.

\begin{figure}[htb]
\centering
\includegraphics[scale=0.48, trim=3cm 3cm 4cm 14cm, clip]{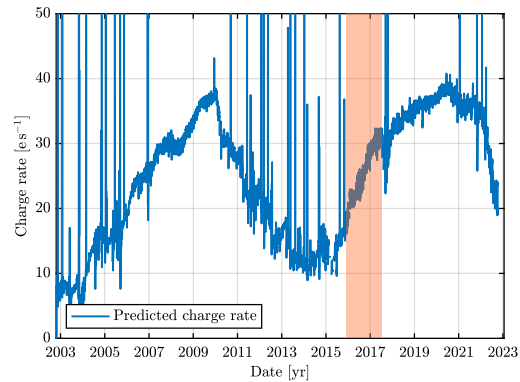}
\caption[Q Timeline.]{\label{fig:Qtimeline} Charging rate of the LISA Pathfinder test masses extrapolated over the 20-year observation range of the INTEGRAL mission by correlation with the IREM count rate as described in the text. The shaded region shows the duration of the LISA Pathfinder mission.}
\end{figure}

Figure~\ref{fig:Qtimeline} shows the resulting estimated test mass charging timeline from 2002-10-20 to 2022-10-10 covering two solar minima and maxima. 
The charging rate over the time of the LPF mission, is indicated by the orange band.  
The minimum charge rate seen is 5-9\,e\,s$^{-1}$ during the solar maximum of 2003 and 10-13\,e\,s$^{-1}$ during the somewhat weaker maximum of 2014. In solar minimum conditions during 2010 and 2020 the charge rate peaks at 36-39\,e\,s$^{-1}$.
The variation in the cosmic ray charging rate is within the ranges that have been assumed in defining the requirements of the LISA mission. The LISA requirements for the allowable charge on the test mass are stricter than was implemented on LPF with a maximum absolute charge of 2.4\,pC equivalent to 70\,mV. Allowing the charge to accumulate from the negative limit, these charge rates would set the discharge frequency in LISA between roughly once per 70 days and once per 9 days.

Although this charging prediction is based on measurements based on LISA Pathfinder and not LISA, given the expected similarity between the test mass and its surrounding electrode housing in the two missions it nonetheless gives a valuable insight into the range of charging rates that can be expected and the rate at which they may change over time. 
Uncertainties in the scaling between charging and RM count rate introduce an uncertainty of up to 10\% as shown in the rates measured throughout the LPF mission in Figure~\ref{fig:QRvsCRMeas}. Electrostatic contributions to the charging rate could also be different in the LISA GRS and could introduce a systematic uncertainty at the level of 20\% as seen in Figure~\ref{Fig:TMChargingACVoltages}. 
A further caveat to apply to these results is that the assumed linear correspondence between the radiation monitor count rates and TM charging has not been established in solar maximum conditions when the energy spectrum of the cosmic ray flux is  different from solar minimum. This could add a systematic error to the lower end of the estimated charge rate range. 

The baseline discharging strategy for LISA is to employ a continuous UV illumination scheme that maintains the test mass potential constant throughout the mission. In the case this approach introduces excess noise however, the fallback scheme would be to discharge the test masses intermittently as was demonstrated successfully in LPF. The estimated range of charging rates over the course of a solar cycle sets the required test-mass discharge frequency. Assuming a 1-hour discharging operation for all LISA test masses simultaneously, and in the pessimistic case that the data during discharging is unusable for science, the projected worst-case discharge frequency would cause an observatory down time of 0.5\%.


\section{Discussion}
\label{discussion}
In this paper we have presented a summary of the test mass charging measurements made on board LISA Pathfinder in order to characterize the charging process. Upon release the test mass potential was consistently negative with respect to the 0\,V reference of the GRS FEE and its amplitude varied from $-512$\,mV to $-12$\,mV. This gives an idea of the triboelectric charging of the TM upon release by the grabbing, positioning and release mechanism and the consequent burden on the control system for electrostatic capture of the TM and the discharge system for its initial neutralization.

We have shown that in both LPF GRS, the test mass charging rate exhibited a strong dependence on the test mass potential due to the the presence of electrons with eV-scale energies emitted from the surfaces of the test mass and surrounding surfaces. The details of the underlying physics of this process are explored in a number of separate publications~~\cite{grimani21a, grimani21b, grimani22,Taioli2023,Wass2022}. Measurements of the electrostatic charge rate dependence indicated that an equilibrium test mass potential existed---at which the net environmental rate of TM charging is zero---between $V_{\textsc{tm}}=600$\,mV and 950\,mV at various stages of the mission. The existence and value of this equilibrium potential means that, if reproduced on LISA, and in the absence of a functioning UV discharge system due to an instrument failure, the TM potential will at least remain stable and bounded. The time constants for the test masses to reach this stable equilibrium are on the scale of 2-3 months. To give an idea of the cost in TM acceleration noise due to coupling of the charge to actuation voltage fluctuations on the GRS electrodes, we note that LPF measurements showed an acceleration noise increase of order 3\,fm\,s$^{-2}$\,Hz$^{-1/2}$ at 0.1\,mHz per TM with $V_{\textsc{tm}}$ = 1\,V,~\cite{Armano2017a} out of a full TM acceleration noise budget of roughly 10\,fm\,s$^{-2}$\,Hz$^{-1/2}$ at that frequency.  

The impact of SEP events on the equilibrium potential produced by environmental charging remains to be studied in detail. During such events, the particle flux can be augmented by several orders of magnitude which will likely drive the test mass to an equilibrium potential rapidly. However, the difference in the energy spectrum of particles responsible for charging due to an SEP event may alter the balance of charging produced by primaries and low-energy secondary electrons and shift the equilibrium away from its steady-state value. 

We have quantified the variations in the test mass charging rate observed over the course of the mission as a function of the on-board radiation monitor record of the galactic cosmic ray flux. The RM count rate provided an accurate analog to the test mass charge rate---especially during flux variations on the scale of days---able to track changes at the 10\% level. 

Due to the short duration of the LPF mission, the energy distribution of the cosmic ray flux is not expected to have changed significantly and no solar energetic particle enhancements were seen. The accuracy of extending the analysis of the count rate dependency outside the conditions of cosmic ray flux approaching minimum in solar activity is therefore uncertain. For a longer duration mission such as LISA, the ability to determine the energy spectrum of the particle flux above 100\,MeV\,nucl$^{-1}$, combined with detailed particle tracking simulations would enhance the ability to correlate charging and particle detection rate. Despite this limitation we have shown that simple particle count-rate data can be used to track short-term variations in the charging rate accurately. 

We have made use of the correlation between charging and the RM count rate and the correlation between the LPF RM count rate and that measured by the IREM TC2 channel to present a 20-year timeline of test mass charging based on historic data. Using this projection, we extrapolate an estimate for the range of test mass charging rates that may be seen in a long-duration mission with a similar GRS such as LISA. The calculated range of 5-39\,e\,s$^{-1}$ is subject to considerable uncertainty at the lower bound which could be resolved by detailed particle tracking simulations with accurate GCR flux models.


\section*{Acknowledgements}
This work has been made possible by the LISA Pathfinder mission, which is part of the space-science programme of the European Space Agency.
The French contribution has been supported by the CNES(Accord Specifique de projet CNES1316634/CNRS103747), the CNRS, the Observatoire de Paris and the University Paris-Diderot. E.Plagnol and H.Inchausp\'e would also like to acknowledge the financial support of the UnivEarthS Labex program at Sorbonne Paris Cit\'e (ANR-10-LABX-0023 and ANR-11-IDEX-0005-02). The Albert-Einstein-Institut acknowledges the support of the German Space Agency, DLR. The work is supported by the Federal Ministry for Economic Affairs and Energy based on a resolution of the German Bundestag (FKZ50OQ0501 and FKZ50OQ1601). The Italian contribution has been
supported by Istituto Nazionale di Fisica Nucleare (INFN) and Agenzia Spaziale Italiana (ASI), Project No. 2017-29-H.1-2020 ``Attivit\`{a} per la fase A della missione LISA''.The Spanish contribution has been supported by contracts AYA2010-15709(MICINN), ESP2013-47637-P, and ESP2015-67234-P(MINECO). M.Nofrarias acknowledges support from Fundacion General CSIC (Programa ComFuturo). F. Rivas acknowledges an FPI contract (MINECO). The Swiss contribution acknowledges the support of the Swiss Space Office (SSO) via the PRODEX Programme of ESA. L. Ferraioli is supported by the Swiss National Science Foundation. The UK groups wish to acknowledge support from the United Kingdom Space Agency (UKSA), the University of Glasgow, the University of Birmingham, Imperial College London, and the Scottish Universities Physics Alliance (SUPA). T. J. Sumner also acknowledges support from the Leverhulme Trust (EM-2019-070\textbackslash4),  J. I. Thorpe and J. Slutsky  acknowledge the support of the U.S. National Aeronautics and Space Administration (NASA).


\bibliographystyle{unsrt2}
\bibliography{paper3}

\end{document}